\documentstyle[12pt]{article}
\newcommand{\lettersection}[1]{\noindent {\it #1} \\}
\newcommand{\be}{\begin{eqnarray}}
\newcommand{\ee}{\end{eqnarray}}
\newcommand{\ba}{\begin{array}}
\newcommand{\ea}{\end{array}}
\newcommand{\bpict}{\begin{picture}}
\newcommand{\epict}{\end{picture}}
\newcommand{\bfig}{\begin{figure}}
\newcommand{\efig}{\end{figure}}
\newcommand{\tr}{{\rm tr}}
\newcommand{\projP}{I \! \! P}
\newcommand{\da}{{\tt a}}
\newcommand{\db}{{\tt b}}

\newcommand{\fa}{{\it a}}
\newcommand{\fb}{{\it b}}

\newcommand{\eh}{{1\over 2}}

\newcommand{\ehti}{\raisebox{.7mm}{\hbox{\tiny 1}}
        \hbox{\tiny /}
        \raisebox{-.7mm}{\hbox{\tiny 2}}}
\newcommand{\dhti}{\raisebox{.7mm}{\hbox{\tiny 3}}
        \hbox{\tiny /}
        \raisebox{-.7mm}{\hbox{\tiny 2}}}
\newcommand{\ehss}{\raisebox{.7mm}{\hbox{\scriptsize 1}}
        \hbox{\scriptsize /}
        \raisebox{-.7mm}{\hbox{\scriptsize 2}}}
\newcommand{\dhss}{\raisebox{.7mm}{\hbox{\scriptsize 3}}
        \hbox{\scriptsize /}
        \raisebox{-.7mm}{\hbox{\scriptsize 2}}}
\newcommand{\partialsl}{\not \! \partial}
\newcommand{\ksl}{\not \! k}
\newcommand{\pslash}{\not \! p}
\newcommand{\qsl}{\not \! q}
\newcommand{\pslashpumu}{{\pslash p_{\mu} \over p^2}}
\newcommand{\intk}{\int {d^4k \over (2\pi)^4}\ }

\newcommand{\intq}{\int {d^4q \over (2\pi)^4}\ }
\newcommand{\intqe}{\int {d^4q_1 \over (2\pi)^4}\ }
\newcommand{\intqz}{\int {d^4q_2 \over (2\pi)^4}\ }
\newcommand{\pqminuse}{{p\over 2}-q_1}
\newcommand{\pqpluse}{{p\over 2}+q_1}
\newcommand{\lambdah}{\raisebox{.7mm}{\hbox{\scriptsize \, $\lambda^A$}}
        \hspace{-.5ex} \hbox{\scriptsize /}
        \raisebox{-.7mm}{\hbox{\scriptsize 2\, }}}
\newcommand{\munu}{_\mu^{\ \nu}}
\newcommand{\gmunu}{g\munu}
\newcommand{\aprojx}{^{(\ehti,0)}\! \projP}
\newcommand{\bprojx}{^{(1,\ehti\,\mid\, \ehti)}\! \projP}
\newcommand{\cprojx}{^{(1,\ehti\,\mid\, \dhti)}\! \projP}
\newcommand{\lprojx}{^{ll}\! \projP}
\newcommand{\tprojx}{^{tt}\! \projP}
\newcommand{\wprojx}{^{w}\! \projP}
\newcommand{\ssprojx}{^{ss}\! \projP}
\newcommand{\llprojx}{^{ll}\! \projP}
\newcommand{\ttprojx}{^{tt}\! \projP}
\newcommand{\slprojx}{^{sl}\! \projP}
\newcommand{\lsprojx}{^{ls}\! \projP}
\newcommand{\stprojx}{^{st}\! \projP}
\newcommand{\tsprojx}{^{ts}\! \projP}
\newcommand{\ltprojx}{^{lt}\! \projP}
\newcommand{\tlprojx}{^{tl}\! \projP}
\newcommand{\lagrangenjla}{{\cal L }
     = \bar q (i\partialsl-m_0)q
      -{g\over 2}\ j^A_{\mu} j^{A\mu} }
\newcommand{\colourcurrent}{j^A_{\mu}
     = \bar q \lambdah \gamma_\mu q  }
\newcommand{\fierztrafoaC}{-{g\over  2}\ j^A_{\mu} j^{A\ \mu}
      = {g\over 3}\, \left(
                     (\bar q \Lambda_{\alpha} q)
                     (\bar q \Lambda^{\alpha} q)
                    +(\bar q      \Gamma_{\alpha} C q^T)
                     (\bar q^T C  \Gamma^{\alpha}   q  )
                     \right)
                  }
\newcommand{\defLambdaqqbar}{\Lambda^{\alpha}
    : = 1_c {\lambda^{\fa}\over 2} O^{\da} }
\newcommand{\defGammaqq}{\Gamma^{\alpha}
    : = {i\epsilon^A\over\sqrt{2}} {\lambda^{\fa}\over 2} O^{\da} }
\newcommand{\indexalpha}{\alpha \equiv (A,\fa,\da)}
\newcommand{\diraco}{O^{\da}
    \in \{i\gamma^5,\,1,\,
          {i\gamma^{\mu}\over\sqrt{2}},\,
          {i\gamma^{\mu}\gamma^5\over\sqrt{2}}
        \} }
\newcommand{\dysonschwingernjla}{
      M_i = m_{0i}
          + {g \over 3} \intk \tr (G_i(k)) }
\newcommand{\quarkpropconstmasstext}{G_{ij} (k)
     = \delta _{ij} G_i (k)
     = \delta _{ij} (\ksl - M_i)^{-1} }
\newcommand{\mesbethesalpimpulsc}{
     \left( - {3\over 2g}{\bf g}_\alpha^{\ \beta}
            - \intk \tr
              [\Lambda_{\alpha} G(k+q/2)
               \Lambda^{\beta}  G(k-q/2)] \right) \phi_\beta(q)
     \vert_{q^2=M_m^2}
     = 0 }
\newcommand{\defdiracalphabeta}{g^{\da\db}
     = \left\{ \ba {llll}
      & \hspace{-2ex} \delta^{\da\db}
         &\hbox{for} &O^\da, O^\db \in \{1,i\gamma^5\}  \\
      & \hspace{-2ex} \delta^{\da\db} g^{\mu\nu}
         &\hbox{for} &O^\da, O^\db \in \{i\gamma^{\mu}         / \sqrt{2},
                                     i\gamma^{\mu}\gamma^5 / \sqrt{2}\}
        \ea \right.  }
\newcommand{\defcfdgalphabeta}{{\bf g}^{\alpha\beta}
     = \delta^{AB} \delta^{\fa\fb} g^{\da\db} }
\newcommand{\defgalphabeta}{\defcfdgalphabeta \ ; \hspace{5ex}
\defdiracalphabeta}
\newcommand{\faddeevL}{
     - \intqz L_\alpha^{\ \gamma}(p,q_1,q_2) \psi_{\gamma}(p,q_2)
     = 0 }
\newcommand{\deffaddeevLGDH}{
     L^{\alpha\gamma}(p,q_1,q_2)
     = i G(\pqpluse) D^{\alpha\beta}(\pqminuse)
       H^{\beta\gamma}(q_1,q_2) }
\newcommand{\impulsH}{
     H^{\beta\gamma}(q_1,q_2)
     = \Gamma^{\gamma} G(-q_1-q_2) \Gamma^{\beta}  }
\newcommand{\diqinvpropimpulsb}{(D^{-1})^{\alpha\beta}(q)
     = - {3\over 4g}{\bf g}^{\alpha\beta}
       - {1\over 2} \intk \tr
         [\Gamma^{\alpha} G(k+q/2)
          \Gamma^{\beta}  G(k-q/2)] }
\newcommand{\staticprop}{
     G(q) = {1\over \qsl - M} \rightarrow - {1\over M} }
\newcommand{\staticfaddeevspinL}{
     \left({\bf g}_\alpha^{\ \gamma} - L_\alpha^{\ \gamma}\right)
     \phi_{\gamma}(p)
     = 0 }
\newcommand{\defstaticL}{
     L^{\alpha\gamma}(p) := \intqe L^{\alpha\gamma}(p,q_1) }
\newcommand{\defstaticphi}{
     \phi^{\gamma}(p) := \intq \psi^{\gamma}(p,q) }
\newcommand{\staticspinLskaxmatrix}{
     \left( \ba {cc}
     1 - (L^{0\, 0})          & \ \         - (L^{0\, 1})^\nu            \\
       - (L^{1\, 0})_\mu      & \ \  \gmunu - (L^{1\, 1})\munu
     \ea \right) }
\newcommand{\staticspinphiskaxvector}{
     \left( \ba {l}
     \phi^0 \\
     \phi^1_\nu
     \ea \right) }
\newcommand{\staticfaddeevspinLskax}{
     \staticspinLskaxmatrix \staticspinphiskaxvector = 0 }
\newcommand{\lorentzdiq}{
  \Bigl( (\ehss\,,0)\oplus(0\,,\ehss) \Bigr) \otimes
  \Bigl( (\ehss\,,0)\oplus(0\,,\ehss) \Bigr)
  = 2\, \Bigl(0\,,0\Bigr)
  \oplus 2\, \Bigl(\ehss\,,\ehss\Bigr)
  \oplus     \Bigl( (1\,,0)\oplus(0\,,1) \Bigr)
        }
\newcommand{\lorentzvectspin}{
  \Bigl(\ehss\,,\ehss\Bigr) \otimes \Bigl( (\ehss\,,0)\oplus(0\,,\ehss) \Bigr)
 = \Bigl( (\ehss\,,0)\oplus(0\,,\ehss) \Bigr)
  \oplus \Bigl( (1\,,\ehss)\oplus(\ehss\,,1) \Bigr)
           }
\newcommand{\sss}{s = \gamma_5}
\newcommand{\einsssP}{1 = s s }
\newcommand{\aprojshortvs}{\aprojx\munu
  = a_\mu a^\nu }
\newcommand{\bprojshortvs}{\bprojx\munu
  = b_\mu b^\nu }
\newcommand{\cprojshortvs}{\cprojx\munu
  = \gmunu - a_\mu a^\nu - b_\mu b^\nu }
\newcommand{\avs}{a_\mu={\gamma_\mu \over 2} }
\newcommand{\bvs}{b_\mu={1\over \sqrt{3}}({\gamma_\mu \over 2} - 2\pslashpumu)
}
\newcommand{\propcprojvsa}{\gamma^\mu\ \cprojx\munu = 0  \ , \ \ \
                       p^\mu\ \cprojx\munu = 0     }
\newcommand{\lprojshortvs}{\lprojx\munu
  =  l_\mu \; l^\nu }
\newcommand{\tprojshortvs}{\tprojx\munu
  =  t_\mu \; t^\nu }
\newcommand{\wprojshortvsPL}{\wprojx\munu
  = \gmunu - l_\mu l^\nu - t_\mu t^\nu \ =\ \cprojx\munu}
\newcommand{\lvsPL}{l_\mu
  = \eh              a_\mu - {\sqrt{3}\over 2} b_\mu
  = \pslashpumu}
\newcommand{\tvsPL}{t_\mu
  = {\sqrt{3}\over 2} a_\mu + \eh              b_\mu
  = {1\over \sqrt{3}} (\gamma_\mu - \pslashpumu)}
\newcommand{\ssprojshort}{\ssprojx   \equiv     1        = s     \; s }
\newcommand{\slprojshort}{\slprojx^\nu  = s     \; l^\nu }
\newcommand{\lsprojshort}{\lsprojx_\mu  = l_\mu \; s     }
\newcommand{\stprojshort}{\stprojx^\nu  = s     \; t^\nu }
\newcommand{\tsprojshort}{\tsprojx_\mu  = t_\mu \; s     }
\newcommand{\ltprojshort}{\ltprojx\munu = l_\mu \; t^\nu }
\newcommand{\tlprojshort}{\tlprojx\munu = t_\mu \; l^\nu }
\newcommand{\defstaticspinsphi}{^s\phi     := s\,      \phi^0}
\newcommand{\defstaticspinlphi}{^t\phi     := t^\nu\,  \phi^1_\nu}
\newcommand{\defstaticspintphi}{^l\phi     := l^\nu\,  \phi^1_\nu}
\newcommand{\defstaticspinwphi}{^w\phi_\mu := \, \wprojx\munu\, \phi^1_\nu}
\newcommand{\propstaticspinwphi}{\gamma^\mu\ \ ^w\phi_\mu =\ 0\ =  p^\mu\ \
^w\phi_\mu }
\newcommand{\staticspindirac}{
     \left(   {\phantom |}^{s_1 s_2}A(p^2)\, \pslash
         -    {\phantom |}^{s_1 s_2}B(p^2)
     \:\right){\phantom |}^{s_2}\phi\ (p)\ =\ 0
     \ ; \ \ \ s_i\in\left\{s,l,t\right\} }
\newcommand{\staticspinrasch}{
     \left(   {\phantom |}^{w}A(p^2)\, \pslash
         -    {\phantom |}^{w}B(p^2)
     \:\right){\phantom |}^{\ w}\phi_\nu(p)\ =\ 0 }
\newcommand{\staticspinehApslB}{
     {\phantom |}^{s_1 s_2}A\, \pslash - {\phantom |}^{s_1 s_2}B
     = \left\{ \ba {llll}
      & \hspace{-2ex}  s_1\    \left(1       - L^{0\, 0}      \right)\ s_2
         &\hbox{if} & s_1=s=s_2\\
      & \hspace{-2ex} -s_1\                   (L^{0\, 1})^\nu        \ s_{2\nu}
         &\hbox{if} & s_1=s,\ s_2 \in \{ l,t \}\\
      & \hspace{-2ex} -s_1^\mu\               (L^{1\, 0})_\mu        \ s_2
         &\hbox{if} & s_1 \in \{ l,t \},\ s_2=s\\
      & \hspace{-2ex}  s_1^\mu\ \left(\gmunu -(L^{1\, 1})\munu\right)\
s_{2\nu}         &\hbox{if} & s_1, s_2 \in \{ l,t \}
        \ea \right.  }
\newcommand{\staticspindhApslB}{
     \left({\phantom |}^w A\, \pslash - {\phantom |}^w B\right)
     {\phantom |}^w \projP_{\bar\mu}^{\ \bar\nu}
     = {\phantom |}^w \projP_{\bar\mu}^{\ \mu}\
       \left(\gmunu -(L^{1\, 1})\munu\right)\
       {\phantom |}^w \projP_{\nu}^{\ \bar\nu} }
\newcommand{\decompaxprop}{
     D^{1,\mu\nu}(k)=
     \left(g^{\mu\nu}-{k^\mu k^\nu \over k^2}\right)  D^{1,trans}(k^2)
     +                {k^\mu k^\nu \over k^2}         D^{1,long} }
\newcommand{\propaxdiq}{
     D^{1,long} = D^{1,trans}(k^2=0) }
\newcommand{\colorstructH}{H^{BC}_{LM}
     \sim {i\epsilon^C_{LK}\over \sqrt{2}} {i\epsilon^B_{KM}\over \sqrt{2}}
     = - \phantom{I}^{1_c}\projP^{BC}_{LM}
       + {1     \over 2} \phantom{I}^{8_c}\projP^{BC}_{LM} }
\begin{document}
%
% BEGIN TITLEPAGE
%
\renewcommand{\thefootnote}{\fnsymbol{footnote}}
\setcounter{footnote}{1}
\begin{center}
\begin{large}
{\bf Octet and Decuplet Baryons in the Hadronized Nambu--Jona-Lasinio
Model with Proper Spin Projection
\footnote {Supported by COSY under contract 41170833} } \\
\end{large}
\vspace{.4cm}
A. Buck, H. Reinhardt \\
Institut f\"ur Theoretische Physik, Universit\"at T\"ubingen \\
Auf der Morgenstelle 14 \\
D--72076 T\"ubingen, Germany
\end{center}
\vspace{.1cm}
\begin{abstract}
\noindent
Octet and Decuplet baryons are described
within the hadronized NJL model as diquark--quark states,
which are bound by quark exchange.
Including scalar and axial--vector diquark correlations,
we project the previously obtained relativistic Faddeev equation
in a Poincar\'e invariant fashion onto good spin,
using a static approximation to the exchanged quark.
The resulting equations
for the spin $\ehss$ octet and the spin $\dhss$ decuplet
are solved numerically.
\end{abstract}
\vspace{.1cm}
%
%
% BEGIN TEXT
%
\renewcommand{\thefootnote}{\arabic{footnote}}
\setcounter{footnote}{0}
\noindent
\lettersection{1.\ Introduction}
Except for some lattice calculations at low energies,
a description of hadrons within QCD is still not feasible.
One therefore resorts to
effective quark models,
which mimic the low energy flavor dynamics of QCD.
In this respect the Nambu-Jona-Lasinio (NJL) model
\cite{NamJon61a}
has proved very successfully in the past.
The success of this model is mainly due to its chiral symmetry,
which almost entirely determines the low energy meson dynamics.
The color--NJL model (see equation (\ref{lagrangenjla}) below)
describes quarks interacting via a local color--octet current interaction,
which for a large number of colors $N_C$
reduces to an attractive quark--antiquark color--singlet interaction
giving rise to the formation of mesons.
The bosonization of this model leads to an effective meson theory
\cite{EbeRei86},
which is in fair agreement with the experimental findings.
Within this effective meson theory
baryons appear as solitons of the meson fields
\cite{ReiWue88},
consistent with Witten's conjecture on large $N_C$--QCD
\cite{Witten79}.
For a finite number of colors
the NJL model contains not only attractive quark-antiquark correlations
but also attractive quark-quark correlation.
In this case the model can be hadronized
i.e. converted into an effective hadron theory,
which contains beside mesons also explicit baryon fields
\cite{Reinhardt90}.
The latter appear as diquark--quark states, which are bound by quark exchange
\footnote{Let us also emphasise that in this approach
it is not necessary that the diquarks are bound.
The diquarks serve only as a convenient building block of the baryons.
This is different in phenomenological diquark--quark descriptions of baryons
\cite{Vogl90,Weiss93}, where elementary diquarks are
considered.}.
Similar investigations have been also performed
in the so--called global color model \cite{Cahill89a},
and in a NJL model with three body forces \cite{Ebert91}.
All these approaches give rise to a relativistic Faddeev equation
for baryons.
\hfill\break\noindent
The relativistic Faddeev equation of the NJL model
has been solved
for spin $\ehss$ baryons in \cite{Buck92} including only scalar diquarks
(which are the counterpart of the pseudoscalar mesons
and hence most strongly bound)
and using the static approximation to the exchanged quark.
Thereby the static approximation was shown to be sufficiently accurate
in \cite{Bentz93a, HuaTjo94}.
Investigations of form factors and magnetic moments
within an additive diquark--quark model
show that the axial--vector diquarks are quite important
even for the spin $\ehss$ baryons \cite{Weiss93}.
The importance of axial--vector diquark correlations is also observed
in the explicit numerical solutions of the Faddeev equation for the nucleon
given in \cite{Bentz93b, Meyer94}.
\hfill\break\noindent
In the present paper we will include the axial--vector diquarks
and solve the Faddeev equation derived in \cite{Reinhardt90}
in the static approximation for both the $s=\ehss$ and $s=\dhss$ hyperons.
For this purpose we will transform the relativistic Faddeev equation
for s-wave bound quark-diquark states
to a system of coupled Dirac and Rarita--Schwinger equations,
respectively.
This Poincar\'e invariant spin projection is much more involved
than the spin algebra in the non-relativistic quark model
or in its relativistic extension given e.g. in \cite{Hussain91}
and has so far not been carried out.
\vspace{0.2cm} \\
\lettersection{2.\ The hadronized NJL model}
Let us briefly outline the relevant results
of the hadronization of the color Nambu--Jona-Lasinio model
(for details see \cite{Reinhardt90}).
The model is defined through the Lagrangian,
\be \lagrangenjla \label{lagrangenjla} \ \ . \ee
Here, $\colourcurrent$, denotes the color octet current,
$\lambda^A\ (A = 1,...\, 8)$ are the $SU_C(3)$ Gell--Mann matrices and
$m_0=diag(m_{0u},m_{0d},m_{0s}$) the current quark mass matrix.
By use of Fierz transformations the interaction can be transformed
into a flavor--singlet quark--antiquark channel
and a flavor--antitriplet quark--quark channel
\be \fierztrafoaC \label{fierztrafoaC} \ \ . \ee
Here, the vertices in the meson and diquark channel are defined by
\be \defLambdaqqbar \ , \ \ \ \defGammaqq \ , \ \ \ \ \indexalpha \ , \ee
where $\epsilon^A_{\ BC}$ stands for the Levi--Civita tensor and
$\ \lambda^\fa\ (\fa = 0,...\, 8)$ are the generators of $U_F(3)$ flavor group.
${O^\da}$ denote the Dirac matrices,
\be \diraco \ , \ee
and C is the charge conjugation matrix.
\hfill\break\noindent
Though the Fierz transformation relates the coupling constants of the
quark--antiquark and the quark--quark channels we will treat them as
independent parameters in the actual calculations.
\hfill\break\noindent
By means of functional integral techniques
the quantum theory defined by the Langrangian (\ref{lagrangenjla})
has been converted into an effective hadron theory \cite{Reinhardt90}
where the baryons are constructed as bound diquark--quark states.
Like in the bosonization of the NJL model \cite{EbeRei86}
the meson fields appear as collective quark--antiquark states .
In the vacuum the scalar meson field develops
an expectation value, $M_{\it i}\ ({\it i} = u,d,s)$,
which signals the spontaneous breaking of chiral symmetry.
This quantity, which represents the constituent quark mass, is defined
by the Schwinger--Dyson equation
\be \dysonschwingernjla \ , \label{dysonschwingernjla} \ee
where $\quarkpropconstmasstext$
is the constituent quark propagator.
The physical mesons represent small amplitude fluctuations $\phi$
around the vacuum expectation value
and the meson masses $M_m$ are defined by the Bethe--Salpeter equation
\be \mesbethesalpimpulsc \label{mesbethesalpimpulsc} \ \ . \ee
Here
${\bf g}^{\alpha\beta}$ is the metric tensor
\be \defgalphabeta \ \ .\ee
\hfill\break\noindent
For the free baryon fields a relativistic Faddeev equation is obtained,
which reads \cite{Reinhardt90}
\be \faddeevL \label{faddeevL} \ , \ee
where,
\be \deffaddeevLGDH \label{deffaddeevLGDH} \ , \ee
contains the two--body quark--quark correlations
through the diquark propagator
\be \diqinvpropimpulsb \label{diqinvpropimpulsb} \ \ . \ee
Furthermore H describes the exchange of a quark,
\be \impulsH \label{impulsH} \ \ . \ee
The relativistic Faddeev equation (8)
has been rederived by more traditional means
in \cite{Bentz93a, HuaTjo94, Bentz93b}
making explicit use of the separability of the NJL interaction.
\hfill\break\noindent
Later on we will employ the static approximation
to the exchanged quark \cite{Buck92},
\be \staticprop \ , \label{staticprop} \ee
which reduces
the integral equation (\ref{faddeevL}) to the algebraic equation
\be \staticfaddeevspinL \label{staticfaddeevspinL} \ , \ee
where
\be \defstaticL \ , \hspace{1cm} \defstaticphi
\label{defstaticL} \label{defstaticphi} \ \ .\ee
\hfill\break\noindent
Below we will include
scalar ($\da=0$) and axial--vector ($\da=1$) diquarks,
which are the counterparts of the pseudoscalar and the vector mesons
and hence the most strongly correlated diquark states of spin 0 and 1.
The Faddeev equation (\ref{staticfaddeevspinL}) then acquires the form
\footnote{Flavor and color indices will usually be supressed in the following.}
\be \staticfaddeevspinLskax \label{staticfaddeevspinLskax} \ , \ee
where $\phi^0$ and $\phi^1_\nu$ denote
the diquark--quark amplitudes (see (\ref{defstaticphi}))
containing a scalar and an axial--vector diquark, respectivly.
\vspace{0.2cm} \\
\lettersection{3.\ Spin projection}
Let us discuss s-wave bound states of three relativistic constituent quarks
from the group theoretical point of view.
Relativistic constituent quarks,
which are described by Dirac spinors
belong to the representation
$(\ehss\,,0)\oplus(0\,,\ehss)$\,
\footnote{As it is well known the representations
of the homogeneous Lorentz group
can be characterized by the quantum numbers $(j_1,j_2)$ of two SU(2) groups.
The representations $(j_1,j_2)$ contain representations
of the Lorentz spin group
with spin
$j=\mid j_1-j_2 \mid \, ,\, ... \, , \, j_1+j_2$, respectively.}
of the homogeneous Lorentz group.
\hfill\break\noindent
The direct product of two quarks is reducible with respect to the decomposition
\be \lorentzdiq \ \ . \nonumber \ee
\hfill\break\noindent
Obviously,
s--wave bound states of a quark and a (pseudo--)scalar diquark
belonging to the $(0,0)$ representation of homogeneous Lorentz group
form baryons in
the spin $\ehss$ representation $(\ehss\,,0)\oplus(0\,,\ehss)$.
Bound states of a quark and an (axial--)vector diquark
belonging to the $(\ehss\,,\ehss)$ representation
are represented by vector spinors.
The vector spinor representation is reducible
according to the homogeneous Lorentz group:
\be \lorentzvectspin \label{lorentzvectspin} \ \ . \ee
Since the irreducible Lorentz representation $(j_1,j_2)$ contains the spins,
$j=\mid j_1-j_2 \mid$, ... , $j_1+j_2$,
the representation $(1\,,\ehss)\oplus(\ehss\,,1)$ contains
both spin $\ehss$ and spin $\dhss$ components.
To project the $(1\,,\ehss)\oplus(\ehss\,,1)$ representation onto good spin
we split the vector spinor representation (\ref{lorentzvectspin}) into
spin $\ehss$ and spin $\dhss$ representations of the Lorentz spin group.
This is achieved by means of the following projectors,
\be \aprojshortvs \ , \hspace{1.cm}
    \bprojshortvs \ , \hspace{1.cm}
    \cprojshortvs \ , \ee
where,
\be \avs \ , \hspace{1.cm} \bvs \ \ . \ee
The spin $\dhss$ projector obeys the Rarita--Schwinger constraints
\be \propcprojvsa \ \ . \ee
The QCD dynamics allows in principle also
for tensor diquark correlations.
The direct product representation
of an antisymmetric tensor diquark and a quark field
can be projected
onto two spin $\ehss$ and two spin $\dhss$ representations.
Thus, s-wave bound states of three constituent quarks
can give rise to eight spin $\ehss$ and four spin $\dhss$ baryon fields
(see also \cite{Carimalo}).
We will restrict ourselves in the following
to scalar and axial--vector diquark correlations.
Then we have three spin $\ehss$ and one spin $\dhss$ baryon fields.
\hfill\break\noindent
Classifying baryons according to the Lorentz spin group
has the unpleasant feature that the projectors
onto the vector spinor (and tensor spinor) subspaces
do no commute with $ \pslash $.
On the other hand
the representations of the Poincar\'e group,
which are characterized by the particle's spin and mass,
do commute with $ \pslash $.
They can be formed
by taking linear transformations of the Lorentz spin group.
For the vector spinor representation (\ref{lorentzvectspin})
the corresponding projectors can be expressed as
\be \lprojshortvs \ , \hspace{1.cm}
    \tprojshortvs \ , \hspace{1.cm}
    \wprojshortvsPL \ , \ee
where,
\be \lvsPL \ , \hspace{1.cm} \tvsPL \ee
and the momentum $p_\mu$  will be fixed later on
by the equation of motion.
The projectors $ \llprojx $, $ \ttprojx $,
the trivial projector,
\be \ssprojshort \ , \ \ \ \sss \ , \ee
and the operators,
\be \slprojshort \ , \hspace{1.cm}
    \stprojshort \ , \hspace{1.cm}
    \ltprojshort \ , \ee
\be \lsprojshort \ , \hspace{1.cm}
    \tsprojshort \ , \hspace{1.cm}
    \tlprojshort \ , \ee
form a complete basis
for all operators acting on spin $\ehss$ s-wave baryons
with scalar and axial--vector diquarks.
In spin $\dhss$ subspace $ \wprojx\munu $ is the identity
\footnote{The operators $\llprojx$, $\ttprojx$, $\ltprojx$, $\tlprojx$
and $\wprojx$ has been firstly derived in \cite{AurUme}
from the Pauli--Lubanski vector in vector spinor representation.}.

\hfill\break\noindent
We now apply this spin projection formalism
to the Faddeev equation (\ref{staticfaddeevspinLskax}).
We define three Dirac spinor fields by the contractions,
\be \defstaticspinsphi \ , \hspace{1.cm}
    \defstaticspinlphi \ , \hspace{1.cm}
    \defstaticspintphi \ , \ee
and a spin $\dhss$ Rarita--Schwinger vector spinor field by
\be \defstaticspinwphi \ \ . \ee
The kernel $L(p)$ defined by
(\ref{deffaddeevLGDH}) and the first equation of (\ref{defstaticL})
is a linear combination of the operators
$^{s_1 s_2}\projP\, ,\ s_i\in\{s,l,t\}$, and $\wprojx\munu$.
Multiplying equation (\ref{staticfaddeevspinLskax}) from the left
by $s$, $l^\mu$, $t^\mu$ and $\wprojx_{\bar\mu}^{\ \mu}$
and using the decompositions
$\einsssP$
and
$ g_\nu^{\ \bar\nu}
= l_\nu l^{\bar\nu} + t_\nu t^{\bar\nu} +\, \wprojx_\nu^{\ \bar\nu}$,
the Faddeev equation reduces to
a system of coupled Dirac equations for the spin $\ehss$ baryon field,
\be \staticspindirac \ \ , \hspace{2.4cm} \label{staticspindirac} \ee
and to a Rarita--Schwinger equation for the spin $\dhss$ baryon field
\be \staticspinrasch \label{staticspinrasch}
    \ ; \ \ \ \propstaticspinwphi \ \ . \ee
Here, the momentum dependent quantities $A$ and $B$ following from equation
(\ref{staticfaddeevspinLskax}) are defined by
\be \staticspinehApslB \label{staticspinehApslB} \ee
and
\be \hspace{-2.0cm} \staticspindhApslB \label{staticspindhApslB} \ . \ee
They involve
the constituent quark propagator
as well as the scalar and the axial--vector diquark propagator,
see equation (\ref{defstaticL}).
\hfill\break\noindent
The axial--vector diquark propagator contains
both a transverse and a longitudinal part,
\be \decompaxprop \label{decompaxprop} \ \ .\ee
Let us emphasize the importance of its longitudinal part.
For Poincar\'e invariant theories
the longitudinal part of a vector propagator is momentum independent
and related to the transverse part by the relation, $\propaxdiq$
\footnote{For a detailed discussion of the Proca equation
in the context of the Poincar\'e group we refer to \cite{AurUme}.}.
The straightforward calculation of the quantities A and B
defined above shows
that it is the longitudinal degree of freedom, which ensures
that the non--kinematic pole of the transverse axial--vector propagator
at $k^2=0$ does not contribute to the Faddeev kernel.
The pole only occurs in the form $(D^{1,trans}(k^2)-D^{1,long})f(k^2)/k^2$,
where $f(k^2)$ is a regular expression for $k^2\rightarrow 0$, and thus
can be removed using the relation (\ref{decompaxprop}).
The longitudinal part of the axial--vector diquark propagator
cannot be descarded in the Faddeev equation because the diquark propagator
is off energy shell.
Neglecting it would give rise to ill--defined integrals.
\vspace{.1cm} \\
Finally some remarks concerning the color and flavor degrees of freedom
are in order.
Color singlet and color octet baryons
decouple since the quark exchange (\ref{impulsH}) has the color structur
(see e.g. \cite{Buck92})
\be \colorstructH \ \ . \ee
Transforming the Faddeev equation (\ref{faddeevL})
into the usual representation of
flavor singlet, $\rho$-- and $\lambda$--type octet and decuplet baryons
(see e.g. \cite{Lichtenberg})
one easily shows that the $SU_F(3)$ multiplets do not decouple
due to the explicit symmetry breaking by the strange quark mass
\footnote{Note that a baryon state with a certain flavor content
(e.g. uud or uds)
need not belong to a specific $SU(3)$--flavor representation.}.
\vspace{0.2cm} \\
\lettersection{4.\ Numerical results}
The dynamical equations (\ref{staticspindirac}) and (\ref{staticspinrasch})
have to be diagonalized in spin flavor space
for the color singlet baryons.
Due to Poincar\'e invariance and flavor conservation
of the interaction (\ref{fierztrafoaC}),
only components with the same spin and with the same flavor content couple.
Furthermore, the Pauli principle forbids
scalar flavor--sextet and axial--vector flavor--antitriplet diquarks
\footnote{For a detailed discussion of the permutation symmetry
of the relativistic three quark bound states we refer
to \cite{Carimalo,Buck95b}.}.
\hfill\break\noindent
In the NJL model arising by Fierz transformation from the Lagrangian
(\ref{lagrangenjla})
the coupling constants
in the (pseudo--)scalar meson channel,
$g_m^0$,
in the (axial--)vector meson channel,
$g_m^1$,
in the (pseudo--)scalar diquark channel,
$ g_d^0$,
and in the (axial--)vector diquark channel,
$g_d^1$,
are related to each other.
Since the local NJL interaction (\ref{fierztrafoaC})
is only a rather crude approximation
to the low energy QCD dynamics,
we prefer to chose different coupling constants in these channels.
This is also tolerated by chiral symmetry.
Then, the model has the following parameters:
The current quark masses, $m_{0u}=m_{0d}$, $m_{0s}$,
the coupling constants, $g_m^0$ and $g_m^1$,\, $g_d^0$,\, $g_d^1$,\,
and the ultraviolett cutoff parameter, $\Lambda$
(see (\ref{mesbethesalpimpulsc}) and (\ref{diqinvpropimpulsb})).
The static approximation (\ref{staticprop}) necessitates the introduction
of a further cut--off for the momentum integral in the reduced Faddeev equation
(see (\ref{staticfaddeevspinL}) and (\ref{defstaticphi})).
In principle, this cut-off is different
from the low energy cut--off of the NJL model
but to reduce the number of parameters we will choose both cut--offs
to be identical.
The numerical results are rather insensitve
to the actual value of the cut--off in the Faddeev equation
in a rather large range.
\hfill\break\noindent
We perform
the momentum integration in the first equation of (\ref{defstaticL})
using a two pole approximation (see \cite{Buck92})
as well as an one pole approximation
to the diquark propagator.
The two pole approximation is in good agreement
within the exact diquark propagator.
Using the one pole approximation scalar and axial--vector diquarks
are considered as Klein--Gordon and Proca particles, respectivly.
\hfill\break\noindent
With a sharp $O(4)$ cut--off
the longitudinal axial--vector diquark propagator
is slightly momentum dependent
and the relation $\propaxdiq$ is violated.
In the present calculations we fix the longitudinal part
of the axial--vector diquark propagator
by means of this relation.
\begin{table} [h]
\centering
\renewcommand{\arraystretch}{1.1}
\begin{tabular}{|l||l||r|r|r|r||r|r|r|r|}
\hline

       & $p$\ \ \      & $\Sigma$\ \ \ & $\Lambda$\ \ \ & $\Xi$\ \ \
       & $\Delta$\ \ \ & $\Sigma^*$\ \ & $\Lambda^*$\ \ & $\Omega$\ \ \  \\
\hline
 I
       & 939 & 1159 & 1165 & 1379 & 1238 & 1440 & 1647 & 1859 \\
\hline
 II
       & 939 & 1117 & 1162 & 1328 & 1238 & 1410 & 1589 & 1777 \\
 III
       & 939 & 1094 & 1166 & 1303 & 1238 & 1411 & 1584 & 1768 \\
\hline
 exp
       &  939 & 1116 & 1193 & 1318 & 1238 & 1385 & 1530 & 1675 \\
\hline
\end{tabular}
\caption[]{\it The masses of the spin $\ehss$ flavor--octet
and the spin $\dhss$ flavor--decuplet baryons:
The results, indicated with (I),
corresponds to the two pole approximation of the diquark propagator,
whereas the results of the rows (II) and (III)
are calculated by means of the one pole approximation.
Constituent masses
$M_u=500$MeV (I,II) and $M_u=430$MeV (III) of the nonstrange quark
are used.
The parameters are fixed as explained in the text.}
\end{table}
\hfill\break\noindent
We fix the parameters,
$g_m^0$, $\Lambda$, $m_{0u}=m_{0d}$ and $m_{0s}$
from the pion decay constant, $f_\pi=93$ MeV,
the pion mass, $m_\pi=138$ MeV,
and the kaon mass, $m_K=496$ MeV, respectively.
This leaves one free parameter, which we choose
(via the gap equation (\ref{dysonschwingernjla}))
to be the non--strange constituent quark mass, $M_u$.
The coupling constants $g_d^0$ and $g_d^1$ are fixed
by the proton and the $\Delta$ masses.
The numerical results for the $\ehss$ and $\dhss$ baryon masses
are presented in table 1.
Using the two pole approximation to the diquark propagator,
see row (I) of table 1,
the ratio of the coupling constants
in the scalar diquark
and pseudoscalar meson sector,
($g_d^{0}/g_m^{0}=1.09$)
deviates only slightly from unity,
the value following from the Fierz transformation
of the color octet NJL model (see (\ref{fierztrafoaC})).
The predicted baryon masses are in fair agreement with the experimental values
except for the $\Omega$ mass,
which is somewhat overestimated.
This is mainly due to the large strange constituent mass
($M_s=719 MeV$ (I,II), $M_s=665 MeV$ (III)).
We are forced to choose large constituent masses
in order to have bound axial--vector diquarks,
since our numerical routine breaks down for unbound diquarks.
Note however that the diquarks need not bound on physical ground.
Allowing also for unbound diquarks we expect a further improvement
of our numerical results.
\vspace{0.2cm} \\
In this paper we have described flavor octet and decuplet baryons
within the hadronized NJL model.
Emphasis is put on a proper Poincar\'e invariant spin projection,
which so far has not been performed in the present diquark--quark picture.
Including scalar and axial--vector diquark correlations
we have derived a system of coupled Dirac equations
for the spin $\ehss$ octet
and of Rarita--Schwinger equations
for the spin $\dhss$ decuplet baryons.
Our numerical calculations show, that the axial--vector diquark admixture
to the spin $\ehss$ baryons is indeed non--negligible.
Finally, let us emphasize that the Poincar\'e invariant spin projection
of the Faddeev equation presented here is not restricted to the NJL dynamic and
can be straightforwardly extended to baryons with orbital excitations.
\vspace{0.2cm} \\
One of us (A.B.) acknowledges discussions with C. Weiss.
\vspace{0.2cm} \\
\lettersection{Note added}
One of the referees brought ref. \cite{HanKre95} to our attention,
which reports on the numerical calculation of hyperon masses
within the hadronized NJL model.
Although the numerical results are similar, our calculations differs
in the following essential point:
As can been seen from the text between equations (4) and (5) of ref.
\cite{HanKre95}, C. Hanhart and S. Krewald neglect
the longitudinal axial--vector diquark correlations.
However, as discussed above, longitudinal axial diquarks are essential in order
to obtain a well defined Faddeev kernel. They are included in the present work.
\end{document}